\documentstyle[aps]{revtex}
\begin{document}
%\draft
%\preprint{HEP/123-qed}
\title{Magnetic ordering in Sr$_2$RuO$_4$ induced by nonmagnetic 
impurities}

\author{M. Minakata}
\address{
Department of Physics, Kyoto University, Kyoto 606-8502, Japan
}
\author{Y. Maeno}
\address{
Department of Physics, Kyoto University, Kyoto 606-8502, Japan\\
CREST, Japan Science and Technology Corporation, Japan
}

\date{\today}
\maketitle
\begin{abstract}
We report unusual effects of nonmagnetic impurities on the spin-triplet
superconductor Sr$_2$RuO$_4$. The substitution of nonmagnetic 
Ti$^{4+}$ for Ru$^{4+}$ 
induces localized-moment magnetism characterized by unexpected Ising 
anisotropy with the easy axis along the interlayer $c$ direction.
Furthermore, for $x\text{(Ti)}\geq 0.03$ magnetic ordering occurs 
in the metallic state with the remnant magnetization along the $c$ axis.
We argue that the localized moments are induced in the Ru$^{4+}$ 
and/or oxygen ions surrounding Ti$^{4+}$ and that the ordering is 
due to their interaction mediated by itinerant Ru-4$d$ elecrtons
with strong spin fluctuations.
\end{abstract}

\twocolumn
\narrowtext

The symmetry of unconventional superconductivity in Sr$_2$RuO$_4$ with 
$T_{\text{c}}=1.5$ K has recently been identified as 
spin triplet \cite{maenoNTR,ishidaNTR,reviews}. 
The initial theoretical insight into the pairing symmetry in Sr$_2$RuO$_4$ 
was based on the similarity of its Fermi-liquid properties to those 
of liquid $^3$He and of the ferromagnetic relative compound SrRuO$_3$ 
\cite{rice}. 
Accordingly, it may seem that ferromagnetic spin fluctuation serves 
as the main origin of spin-triplet pairing. The investigation of 
magnetic fluctuation spectrum by inelastic neutron scattering \cite{sidis}, 
however, revealed a strong susceptibility peak at incommensurate 
wave number $q = Q_{\text{ic}} \cong (2\pi/3a, 2\pi/3a, 0)$, 
which is attributable to 
nesting between two of the three Fermi-surface sheets \cite{mazin}. 
Nevertheless, there is an important second component in the spin 
fluctuation spectrum: strongly enhanced paramagnetism extending 
in a wide $q$-range, which is attributable to the correlation in 
the other Fermi-surface sheet \cite{mukuda}. Thus the roles of these 
band-specific spin fluctuations in the superconducting symmetry 
and mechanism are a subject of active theoretical investigations 
\cite{sato,nomura}. 

In addition to metallic SrRuO$_3$ with ferromagnetic ordering below 
$T_{\text{c}} = 160$ K \cite{SrRuO3}, another relative compound 
Sr$_3$Ru$_2$O$_7$ has recently 
been shown to order ferromagnetically below $T_{\text{c}} = 100$ K under 
pressure \cite{Sr3Ru2O7}. Therefore, it may be possible to induce magnetic 
ordering also in Sr$_2$RuO$_4$ by appropriate modification of the 
compound. In fact, low-temperature transport properties of 
Sr$_2$RuO$_4$ under pressure suggest an important change in the 
electronic state above 1.5 GPa , although its details have not 
been clarified at present \cite{yoshida}. The characterization of the 
magnetic ordering, 
if induced, will add important information on the spin fluctuation, 
which is believed to be responsible for the pairing in Sr$_2$RuO$_4$.

Impurity ions substituting Cu have extensively been used as 
powerful probes to reveal the unconventional properties of 
high-$T_{\text{c}}$ cuprates \cite{maenoCu}. In particular, nonmagnetic 
Zn$^{2+}$ is 
found to induce local moments and suppress $T_{\text{c}}$ severely 
\cite{xiao,CuNMR,fukuzumi}. Here we report a corresponding remarkable 
effect of 
nonmagnetic impurities for Ru in Sr$_2$RuO$_4$. We select Ti$^{4+}$ 
$(3d^0)$, 
a closed-shell ion without $d$ electrons, as a nonmagnetic impurity
because it has the same oxidation number and coordination 
number as Ru$^{4+}$ $(4d^4)$, as well as a very similar ionic radius. 
In fact, the compound Sr$_2$TiO$_4$ is isostructural to Sr$_2$RuO$_4$ with 
very similar lattice parameters: the basal parameter $a$ is longer 
by merely 0.4\% and the interlayer parameter $c$ is shorter by 1.0\% 
than those of Sr$_2$RuO$_4$. We found that the nonmagnetic impurity 
induces localized moment characterized by strong Ising anisotropy 
with the easy axis parallel to the interlayer $c$ axis. 
Furthermore, we found that the magnetic ordering is induced 
while retaining the metallic state. 

We used single crystals of Sr$_2$Ru$_{1-x}$Ti$_x$O$_4$ 
$(0 < x < 0.25)$ grown by a floating zone method with an infrared 
image furnace (NEC Machinery, model SC-E15HD). 
To prepare a feed rod to be melted for the crystal growth, a mixture 
of SrCO$_3$, RuO$_2$ and TiO$_2$ powders with the molar ratio of 
$2:(1.15 - x):x$ were reacted at $1250^\circ\text{C}$ 
for 24 to 48 hours. For higher $x$, Sr$_2$TiO$_4$ was first synthsized 
and mixed with SrCO$_3$ and RuO$_2$.
The excess Ru is to compensate for the active sublimation of 
RuO$_2$ during the crystal growth. Single phase of K$_2$NiF$_4$-type 
tetragonal cell was confirmed by X-ray diffraction measurements 
on powdered crystals. The lattice parameters $a$ increases and 
$c$ decreases systematically with $x$ at the rate of $da/dx = 0.15\%$
and $dc/dx = -1.9\%$ for $0 \leq x \leq 0.09$, consistent with the 
expected tendency for the Ru-Ti substitution. The high-resolution
electron probe microanalysis (EPMA) revealed that the analyzed concentration 
$x_{\text{a}}$ is related with the nominal $x$ by $x_{\text{a}} = 0.95 x$ 
for $x$ up to 0.09 \cite{EPMA}. 
All these clearly indicate that Ti is well substituted for Ru 
in the structure of Sr$_2$RuO$_4$. This is in sharp contrast with 
the Al doping case, in which the analyzed concentration is 
always much lower than the nominal one. We deduce that the trivalent 
Al ions are mainly introduced as lattice defects or interstitials, 
rather than directly replacing Ru, and their magnetic properties 
are quite different from those of Ti ions \cite{Al}. Nevertheless both 
kinds of impurities cause extremely strong depairing: with $x\text{(Ti)} = 
0.001$, the lowest $x$ investigated in the present study, the 
superconductivity is already suppressed below 0.3 K, consistent 
with extreme sensitivity of $T_{\text{c}}$ to impurities and defects 
\cite{Al,defect}.

The DC magnetization and AC magnetic susceptibility were measured 
using a SQUID magnetometer (Quantum Design, model MPMS).
Fig. 1 shows DC magnetization $M/H$ at $H = 10$ kOe (a) parallel 
to the $c$ axis and (b) parallel to the $ab$ plane. The normal state 
of pure Sr$_2$RuO$_4$ shows almost isotropic magnetization with weak 
temperature dependence, characterized by enhanced Pauli 
paramagnetism due to hybridized Ru$^{4+}$ and O$^{2-}$ 
electrons \cite{FL}. In contrast, insulating Sr$_2$TiO$_4$ shows 
a negative value at high temperatures representing the diamagnetism 
of the ion cores. At low temperatures, a weak Curie term 
attributable to impurity ions is visible. When 
Ti$^{4+}$ is introduced in Sr$_2$RuO$_4$, however, the magnetization 
becomes greater than those of both end members. It shows a 
growing Curie-Weiss-like term with increasing $x$, indicating 
that the localized moments are induced by the substitution of 
nonmagnetic Ti$^{4+}$ for Ru$^{4+}$. 
The effective magnetic moment $p_{\text{eff}}$ estimated from the 
Curie-Weiss fitting between 150 and 300 K is somewhat 
smaller than 0.5 $\mu_{\text{B}}/\text{Ti}$ for $0 < x < 0.25$. 
As shown in Fig. 1, the 
induced moment exhibits strong Ising anisotropy with the easy 
axis along the $c$ axis. This anisotropy was quite unexpected 
considering that the magnetization of Sr$_2$RuO$_4$ and 
Ca$_{2-x}$Sr$_x$RuO$_4$ 
with $x$ close to 2 is nearly isotropic \cite{CSRO} and that the moments 
in the antiferromagnetic ground state of the relative Mott 
insulator Ca$_2$RuO$_4$ are lying in the $ab$ plane \cite{Ca2RuO4}.

As the main feature in Fig. 1 (a), magnetic ordering is 
induced for $x \geq 0.03$. With $H$ parallel to the $c$ axis hysteresis 
occurs between zero-field-cooled (ZFC) and field-cooled (FC) 
magnetization for $x \geq 0.03$, and a peak appears in the ZFC 
magnetization at $T_{\text{p}}$. 
To confirm the onset of magnetic order, we performed $M-H$ sequences 
with the results shown in Fig. 2. 
At each temperatue, the crystal ($x$ = 0.09) was first
cooled from 20 K in zero field (open circles).
The field in then increased to 50 kOe (triangles) and subsequently 
decreased to zero again (closed circles).
The remnant magnetization obtained in this sequence 
clearly disappears above $T_{\text{mag}}$ = 14.5 K, 
indicating the magnetic ordering below this temperature. 
Thus, although $T_{\text{p}}$ = 12.5 K somewhat underestimates 
the ordering temperature, it serves as a good measure of it. 
As shown in Fig. 3, $T_{\text{p}}$ increases linearly with $x$ for 
$0.03 \leq x \leq 0.09$ and tend to saturate beyond $x \simeq 0.1$. 

With $H$ parallel to the $ab$ plane, there is no corresponding 
reduction or hysteresis in the $M/H$-vs-$T$ curves. Furthermore, 
magnetization with $H$//$ab$ exhibits no in-plane anisotropy 
at $1.8 - 300$ K for $x = 0.09$, indicating the absence of an easy 
axis in the plane. 
Therefore, we conclude that the moment 
points in the $c$ direction in the ordered state. 
The moments resides most probably on Ru$^{4+}$ ions, 
which surround randomly distributed Ti$^{4+}$. But they may partially resides 
on the neighboring O$^{2-}$ ions as well.
As we will show below, the in-plane conduction is maintained 
in the ordered phase. Thus the magnetism of this system is 
consistent with Ising spin glass formed by RKKY-type interaction 
among localized moments. 

A very slow relaxation of the remnant magnetization, characteristic of 
the spin-glass formation, is indeed observed in the ordered state. 
Moreover, the field and frequency dependence of the ordering is also 
consistent with the spin-glass formation. The magnetization 
in a wider range of fields (10 Oe $<H<$ 50 kOe) shows that 
with decreasing the field the peak of ZFC magnetization becomes 
sharper and $T_{\text{p}}$ increases. The AC susceptibility also shows 
the Ising anisotropy and weak frequency dependence 
(1 Hz - 100 Hz) of the peak temperature.  

In contrast with the clear Ising anisotropy at low temperatures, 
the increase in the magnetization at 300 K is precisely isotropic: 
$d(M/H)/dx = 1.05 \times 10^{-3}$ emu/mol for 
both $H$//$c$ and $H$//$ab$. 
Since the number of $d$ electrons per mol decreases with increasing $x$, 
this indicates that the Pauli component arising from itinerant 
electrons is also enhanced with $x$, implying an approach to Fermi 
surface instability.

Electrical resistivity $\rho(T)$ was measured down to 4.2 K by 
a standard DC four-probe method. As shown in Fig. 4 (a), the 
interlayer resistivity $\rho_{c}$ of pure Sr$_2$RuO$_4$ shows metallic 
behavior below 130 K and nonmetallic $d\rho_{c}/dT < 0$ at higher 
temperatures. The metal-nonmetal crossover temperature 
(at which $d\rho_{c}/dT = 0$) decreases with increasing $x$. 
A low temperature upturn in $\rho_{c}$ appears for $x > 0.04$, 
and for $x > 0.05$ $\rho_{c}$ shows nonmetallic behavior at all 
temperatures below 300 K. The appearance of the low-temperature 
upturn corresponds to the value of $\rho_{c}$ exceeding 30 - 40 m$\Omega$cm. 
In contrast, the in-plane resistivity $\rho_{ab}$ retains metallic 
conduction at high temperatures at least up to $x = 0.09$. 
Figure 4 (b) shows the $x$ dependence of $\rho_{ab}(T)$ which suggests that 
Ti impurities can be treated as strong potential scatterers, 
while they do not seriously affect the inelastic part. 
Metal-insulator transition is expected to occur if the universal 
value to sustain two-dimensional metallic conduction 
$h/4e^{2} = 6.5$ k$\Omega$/square
per RuO$_{2}$ plane is exceeded. 
This value corresponds 
to $\rho_{ab}(0) \cong 400$ $\mu\Omega$cm 
for Sr$_2$RuO$_4$, and is not reached 
for $x \leq 0.09$. The temperature of 
the minimum in $\rho_{ab}$ does not correspond to the magnetic 
ordering.
Rather, the low-temperature upturn, clearly seen for 
$x = 0.09$, is attributable to weak localization caused by 
randomness introduced by Ti impurities. 

Let us discuss now why the moments induced by Ti$^{4+}$ exhibit 
strong Ising anisotropy. Since the Pauli component remains 
isotropic, the induced Curie-Weiss component is ascribable 
to the local moments. 
For a localized state, the cluster calculation of the atomic 
orbits of RuO$_6$ indicates that the orbital moment align along the 
$c$ axis if RuO$_6$ octahedron is undistorted or elongated along the $c$ 
direction, while it is in the plane if the octahedron is 
flattened \cite{mizokawa}. In the presence of the interaction of the 
atomic orbits with spins, this explains why the moments in the 
Mott insulator Ca$_2$RuO$_4$ with flattened octahedra lie in the $ab$ plane. 
Since the octahedron is elongated in the present system, the Ising 
anisotropy along the $c$ axis is reasonable as long as the localized 
picture applies.

The magnitude of the induced local moment, 
 $p_{\text{eff}} \lesssim 0.5$ $\mu_{\text{B}}$ per Ti, 
is substantially smaller than the $2.82$ $\mu_{\text{B}}$ expected for the 
localized moment of Ru$^{4+}$ in the low-spin configuration with $S = 1$.  
Besides, there are four Ru sites 
adjacent to each Ti in the RuO$_2$ plane. Such small moment is also 
induced by Zn$^{2+}$ in the cuprates: $\sim 0.8 \mu_{\text{B}}$ 
in the underdoped region and $\sim 0.2 \mu_{\text{B}}$ 
in the highly doped region \cite{fukuzumi}. 
(The moment induced around Zn is known to be staggered \cite{CuNMR}.) 
These small moments may be due to the partially itinerant character 
of the $d$ electrons of Ru and Cu. In the present case, 
one of the three t$_{\text{2g}}$ orbitals of the 
Ru$^{4+}$ ion in the Ru-O-Ti sites does not involve bonding to the Ti site. 
Thus, such orbital should be relatively unaffected by the absence of the 
$d$ electron at the Ti site and retain itinerant character.
NMR study on Ti substituted Sr$_2$RuO$_4$ may be able to clarify 
whether the local moment is confined only in the Ru adjacent 
to Ti, or involves oxygen as well as more Ru sites. 

In the cuprates, magnetic ions such as Ni$^{2+}$ are known to exhibit 
smaller moments as well as weaker depairing effect than 
nonmagnetic ions \cite{xiao}. A suitable magnetic impurity ion for 
Sr$_2$RuO$_4$ is Ir$^{4+}$ $(5d^{5})$. According to the result on 
polycrystalline Sr$_2$(Ru, Ir)O$_4$, the Pauli component of the 
susceptibility decreases with slight Ir doping and the Curie 
moment is induced with $p_{\text{eff}} \sim 0.3$ $\mu_{\text{B}}/\text{f.u.}$
at $x({\text{Ir}}) = 0.1$, 
corresponding to a large value: 
$\sim 3$ $\mu_{\text{B}}/\text{Ir}$ \cite{cava}. 
Considering the strong anisotropy 
found with Ti substitution, it is well worth repeating the investigation 
with single crystals.

Concerning the origin of the magnetic ordering,
we discussed why it is ascribable to the spin glass formation 
of the induced moments by 
RKKY-type interaction. Here, it should be noted that the itinerant 
electron system mediating the interaction has a strongly $q$-dependent 
susceptibility, especially a strong peak at $q = Q_{\text{ic}}$. 
A recent tight 
binding calculation including the spin-orbit interaction suggests that 
the susceptibility at $Q_{\text{ic}}$ has an enhanced component 
along the $c$ axis \cite{ng}. 
The characteristic length corresponding to $Q_{\text{ic}}$ is 
$\lambda_{\text{ic}} \cong 3 a/\sqrt{2} = 0.82$ nm. 
This is to be compared with 
the mean Ti-Ti distance $d_{ab}$ within the plane, $d_{ab} = a/\sqrt{x}$ 
on the assumption of random distribution of Ti ions. 
The threshold concentration for the magnetic ordering, $x_{\rm c} 
\cong 0.025$, 
corresponds to $d = 6.3 a$, three times longer than $\lambda_{\text{ic}}$. 
We have not observed any indication that the magnetic phase becomes 
particularly stable at a special concentration, possibly related to 
$\lambda_{\text{ic}}$. 
We also note that the incommensurate antiferromagnetic ordering alone 
cannot explain the remnant moment, because the observed tetragonal 
structure (confirmed by neutron scattering at low temperatures 
\cite{braden}) does not allow a ferromagnetic component associated with 
spin canting.
Nevertheless, it is quite important to investigate how 
the spin fluctuation at $Q_{\rm ic}$ is modified by the Ti substitution 
and by the resulting magnetic ordering at low temperatures. 
For this purpose, neutron scattering study on Ti-substituted 
single crystals of Sr$_2$RuO$_4$ is under progress \cite{braden}.

In conclusion, we have shown that nonmagnetic impurities on the Ru site
serve as a very effective probe of spin fluctuation in Sr$_2$RuO$_4$. 
Although the nonmagnetic impurities in both Sr$_2$RuO$_4$ and 
high-$T_{\text{c}}$ cuprates induce local moments and severely 
suppress $T_{\text{c}}$, 
the anisotropy of the induced moments as well as their interaction 
appear substantially different, and results in the magnetic ordering 
in Sr$_2$RuO$_4$. 

%\begin{acknowledgments}
The authors acknowledge A.P. Mackenzie and M. Braden 
for important collaborations and 
N. Kikugawa and T. Iwamoto for their contribution. 
They are grateful to T. Ishiguro for his support, 
to Z.Q. Mao, S. NishiZaki, S. Nakatsuji, H. Fukazawa,
 and K. Deguchi for their technical support and useful 
discussions, and to M. Sigrist, T. Mizokawa, K. Yamada, 
and K. Ishida for valuable discussions.
%\end{acknowledgments}

\begin{figure}
\caption{Temperature dependence of the magnetization of 
Sr$_2$Ru$_{1-x}$Ti$_x$O$_4$  
with the applied field (a) parallel to the $c$ axis and (b) parallel to 
the $ab$ plane. ZFC and FC indicate zero-field-cooled and 
field-cooled data. Note that the scales of the figures differ
by a factor of two.}
%\label{autonum}
\end{figure}

\begin{figure}
\caption{Temperature dependence of the remanent magnetization of 
Sr$_2$Ru$_{1-x}$Ti$_x$O$_4$ ($x$ = 0.09) 
with the applied field parallel to the $c$ axis.
After zero-field-cooling at each temperature (open circles),
the field was increased to 50 kOe (triangles).
The remnant magnetization was measured two minutes after the field was 
decreased to zero (closed circles). 
A small, temperature independent contribution of c.a. 0.032 
$\mu_{\text{B}}/\text{Ti}$, identified as the contribution from 
SrRuO$_3$ impurity, has beeen subtracted.}
%\label{autonum}
\end{figure}

\begin{figure}
\caption{Phase diagram of Sr$_2$Ru$_{1-x}$Ti$_x$O$_4$ 
with the applied field of $\mu_{0}H = 10$ kOe. 
The open circles indicate that no peak of ZFC magnetization is seen 
at $T > 1.8$ K. The dotted line is a guide to the eye.}
%\label{autonum}
\end{figure}

\begin{figure}
\caption{Temperature dependence of (a) the interlayer and (b) in-plane 
resistivity of Sr$_2$Ru$_{1-x}$Ti$_x$O$_4$.}
%\label{autonum}
\end{figure} 

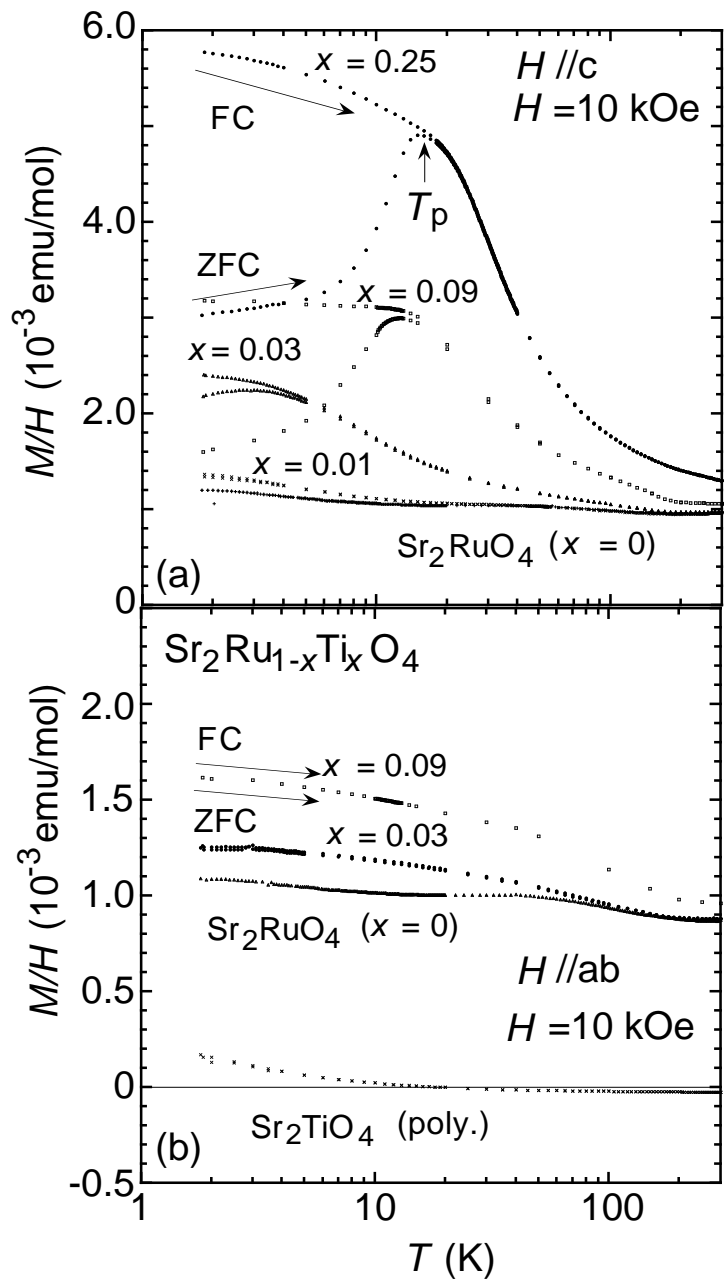
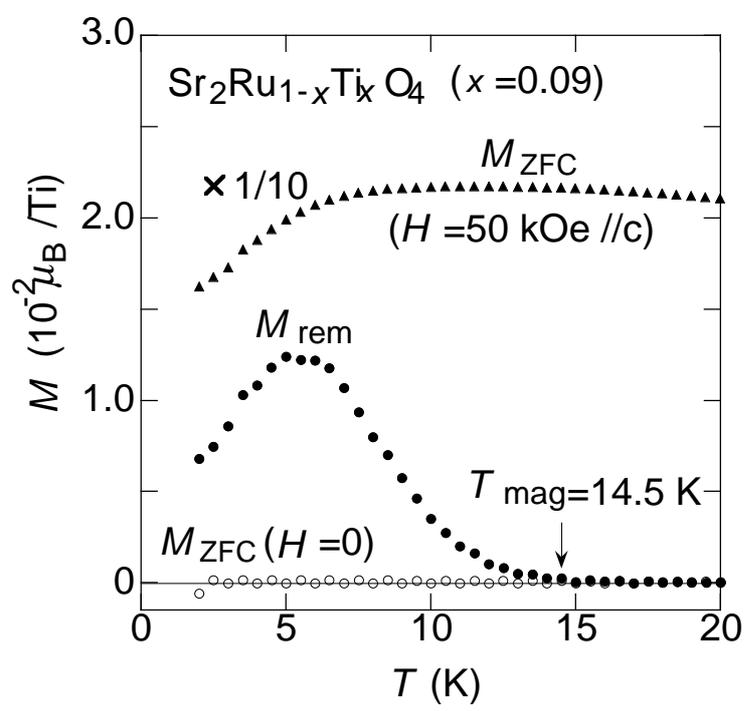
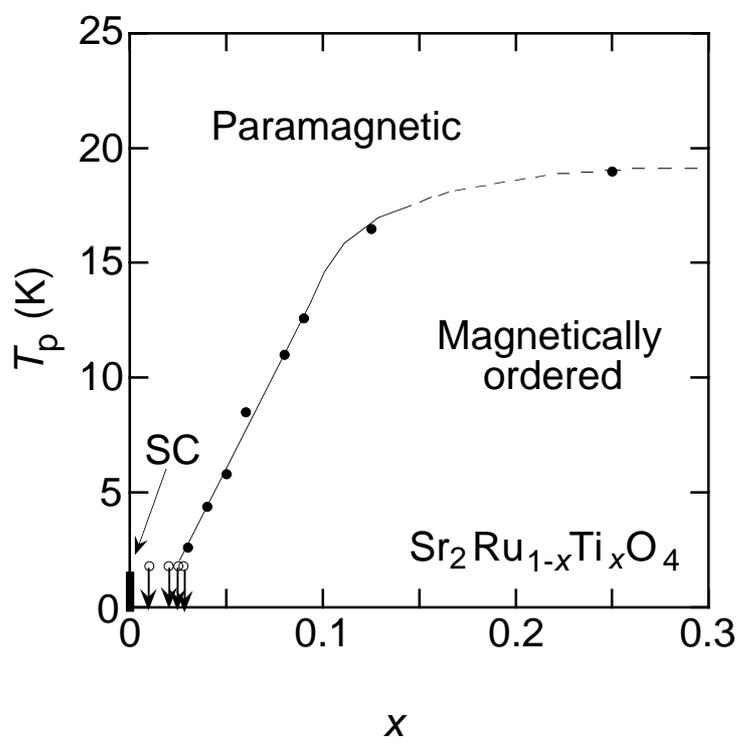
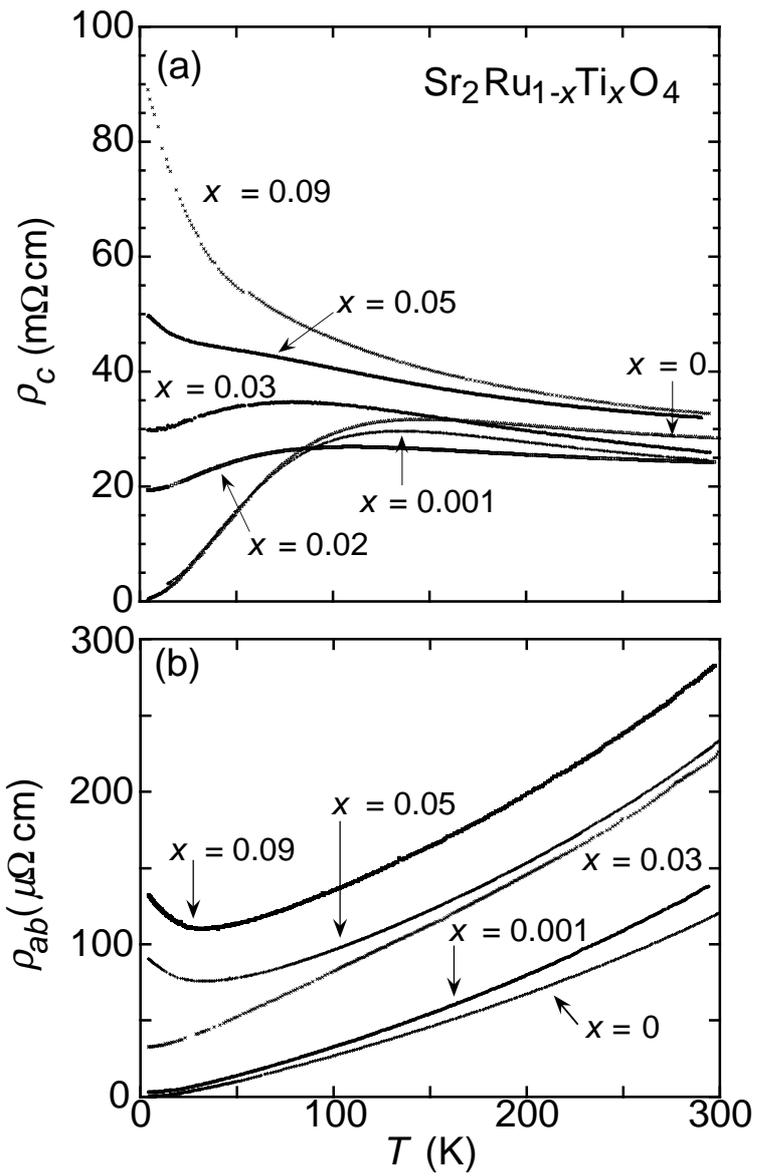
\end{document}